\documentclass[acus]{JAC2003}

\usepackage{graphicx}
\usepackage{booktabs}

\begin{document}
\title{OPTIMIZING INFRASTRUCTURE FOR SOFTWARE TESTING USING VIRTUALIZATION}
\author{O. Khalid, A. Shaikh, B. Copy. CERN, Geneva, Switzerland}
\maketitle

\begin{abstract}
 Virtualization technology and cloud computing have brought a paradigm shift in the way we utilize, deploy and manage computer resources. They allow fast deployment of multiple operating system as containers on physical machines which can be either discarded after use or checkpointed for later re-deployment. At European Organization for Nuclear Research (CERN), we have been using virtualization technology to quickly setup virtual machines for our developers with pre-configured software to enable them to quickly test/deploy a new version of a software patch for a given application. This paper reports both on the techniques that have been used to setup a private cloud on a commodity hardware and also presents the optimization techniques we used to remove deployment specific performance bottlenecks.
\end{abstract}

\section{INTRODUCTION AND BACKGROUND}
This paper reports our work to evaluate emerging software technologies such as virtualization and cloud computing for control system applications especially for small teams to quickly setup test environments for development and testing. Virtualization is a software layer that runs on the underlying hardware, and enables system administrators to run multiple operating systems as isolated applications or precisely speaking as virtual machines (VM). The technology have been around since 60's when IBM first developed it to enable users to share mainframe for their applications in an isolated way.

In recent years, virtualization technology have matured to provide bare-metal performance for virtual machines and have been increasingly deployed at large scale (from clusters to data centers) using cloud computing technology to optimize the utilization of the physical infrastructure in an elastic and flexible manner. According to National Institute of Standards and Technology (NIST), Cloud Computing is `` a model for enabling ubiquitous, convenient, on-demand network access to a shared pool of configurable computing resources (e.g., networks, servers, storage, applications, and services) that can be rapidly provisioned and released with minimal management effort or service provider's interaction" \cite{nist}.

NIST identifies two key characteristics of cloud computing that differentiates it from other ways of organizing and accessing computing infrastructures. First is \textit{on-demand self service} that enables a consumer/user to unilaterally provision computing capabilities on demand without requiring human interaction with the service provider. Second feature of cloud computing is \textit{resource pooling} of computing capabilities by the provider to serve multiple user-communities from the same physical infrastructure creating location independence where the user has no knowledge or control over the provided resources. The service model we have opted to deploy is \textit{Infrastructure as a Service} (IaaS) to provide our user community with a capability to provision computing, storage and networking to run any operating system or application. 

The key motivation to opt for a private cloud has been the way we use the infrastructure. Our user community includes developers, testers and application deployers who need to provision machines very quickly on-demand to test, patch and validate a given configuration for CERN's control system applications. Virtualized infrastructure along side with cloud management software enabled our users to request new machines on-demand and release them after their testing was complete.

\section{IMPLEMENTATION}

\begin{figure*}[tb]
    \centering
    \includegraphics*[width=170mm, height=65mm]{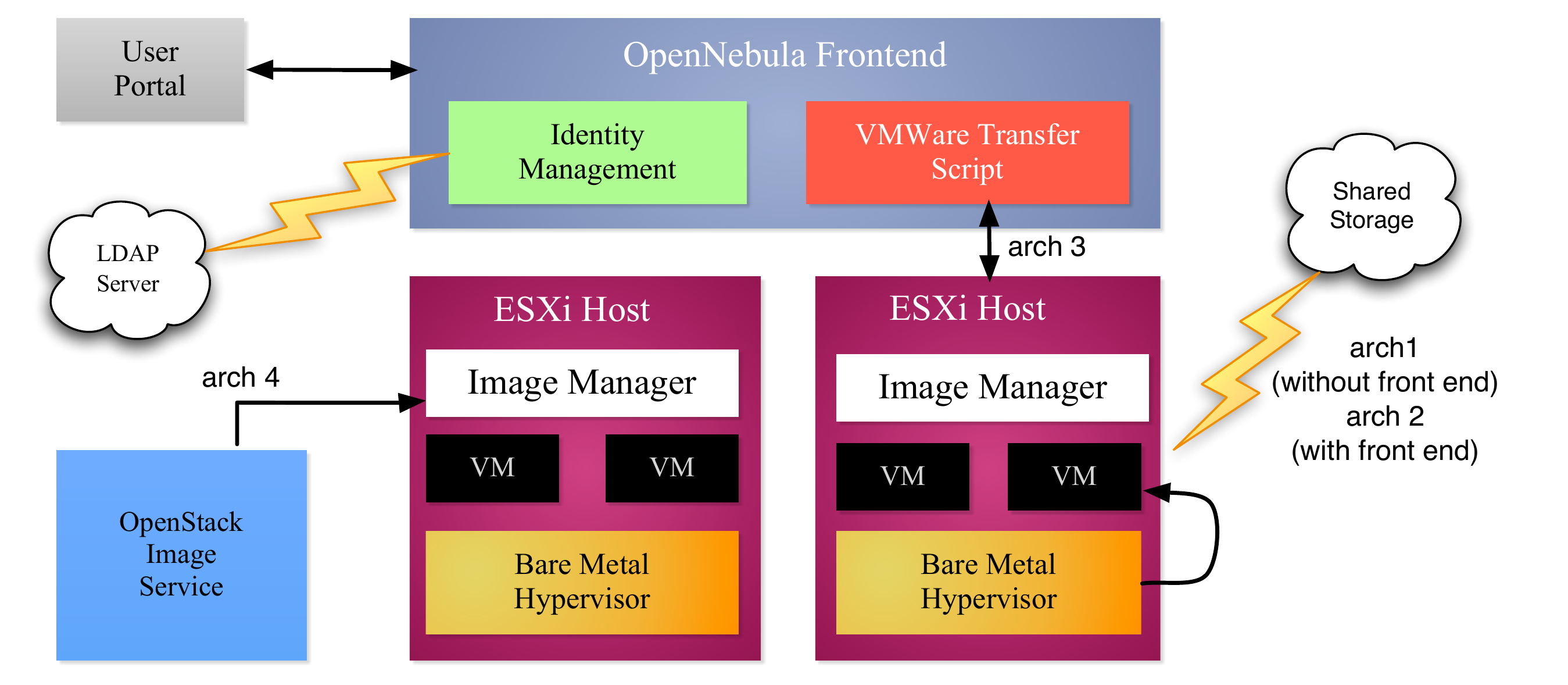}
    \caption{The cloud deployment architecture based on OpenNebula, OpenStack and VMWare ESXi software.}
    \label{arch}
\end{figure*}
The hardware we use for our experimentation is HP Proliant 380 G4 machines with 8GB of memory, 500 GByte of disk and connected with gigabit ethernet. Five servers were running VMWare ESXi bare-metal hypervisor to provide virtualization capabilities \cite{esxi}. We also evaluated Xen hypervisor \cite{xen} with Eucalyptus \cite{euca} cloud but given our requirements for Windows VMs, we opted for VMWare ESXi. OpenNebula Professional (Pro) was used as cloud front-end to manage ESXi nodes and to provide users with an access portal \cite{opennebula}. Number of deployment configurations were tested and their performance was benchmarked. The configuration we tested for our experimentation are the following as show in Fig. \ref{arch}:

\begin {itemize}
\item Central storage without front end (\textit{$arch_1$} ): a shared storage and OpenNebula Pro runs on two different servers. All VM's images reside on shared storage all the time. 
\item Central storage with front end (\textit{$arch_2$} ): a shared storage, using network filesystem (NFS), shares the same server with OpenNebula front end . All VM images reside on shared storage all the time.
\item Distributed storage remote copy (\textit{$arch_3$} ): VM images are deployed to each ESXi node at deployment time, and copied using Secure Shell (SSH) protocol by front end's VMWare transfer driver.
\item Distributed storage local copy (\textit{$arch_4$}): VM images are managed by an image manager service which downloads images pre-emptively on all ESXi nodes. Front end runs on a separate server and setup VM using locally cached images.
\end {itemize}

Each of the deployment configuration has its advantages and disadvantages. \textit{$arch_1$} and \textit{$arch_2$} are using a shared storage model where all VM's are setup on a central storage. When a VM request is sent to the front end, it clones an existing template image and sets it up on the central storage. Then it communicates the memory/networking configuration to the ESXi server, and pointing the location of the VM image. The advantage of these two architectural configuration is that it simplifies the management of template images as all of the virtual machine data is stored on the central server. The disadvantage of this approach is that incase of a disk failure on the central storage; all the VM's will loose data. And secondly, the system performance can be seriously degraded if shared storage is not high performance and doesn't has high-bandwidth connectivity with ESXi nodes. Central storage becomes the performance bottleneck for these approaches.

\textit{$arch_3$} and \textit{$arch_4$} tries to overcome this shortcoming by using all available diskspace on the ESXi servers. The challenge here is how to clone and maintain VM images at run time and to refresh them when they get updated. \textit{$arch_3$} resolves both of these challenges by copying the VM images at request time to the target node (using VMWare transfer script add-on from OpenNebula Pro software), and when the VM is shut then the image is removed from the node. For each new request, a new copy of the template image is sent over the network to the target node. Despite its advantages, network bandwidth and ability of the ESXi nodes to make copies of the template images becomes the bottleneck. \textit{$arch_4$} is our optimization strategy where we implement an external image manager service that maintains and synchronize a local copy of each template image on each ESXi node using OpenStack's Image and Registry service called Glance \cite{openstack}. This approach resolves both storage and network bandwidth issues. 

Finally, we empirically tested all architectures to answer the following questions:

\begin {itemize}
\item How quickly the system can deploy a given number of virtual machines?
\item Which storage architecture (shared or distributed) will deliver optimal performance?
\item What will be average wait-time for deploying a virtual machine?
\end {itemize}

\subsection{Contextualization}

One of the major challenge of deploying windows virtual machines is contextualizing them for a specific environment at deployment time e.g. a windows virtual machine getting a public IP address in CERN network and joining the domain to allow CERN applications to be deployed on it. This requires the VM to part of the public network at the deployment time, and be able to join the domain.

At CERN, network access is controlled by pre-registered list of authorized network interfaces. A list of virtual Machine Access Card (MAC) addresses are pre-registered in the network database. A new VM is deployed with one of the available MAC address. Once its get into a running stage, it gets connected to the network.

Next stage is to configure the machine to acquire the new machine name corresponding to the virtual MAC address. It's simpler to configure for Linux VM's as compared to Microsoft Windows XP VMs. For both we have adopted similar approaches to automate the contextualization process. Each VM is configured to launch a script at boot time that gets its MAC address and compare against a list of registered names (either using a local file or remotely access file). The matching name is updated in the VM and it's rebooted.

\begin{figure}[htb]
   \centering
   \includegraphics*[width=85mm]{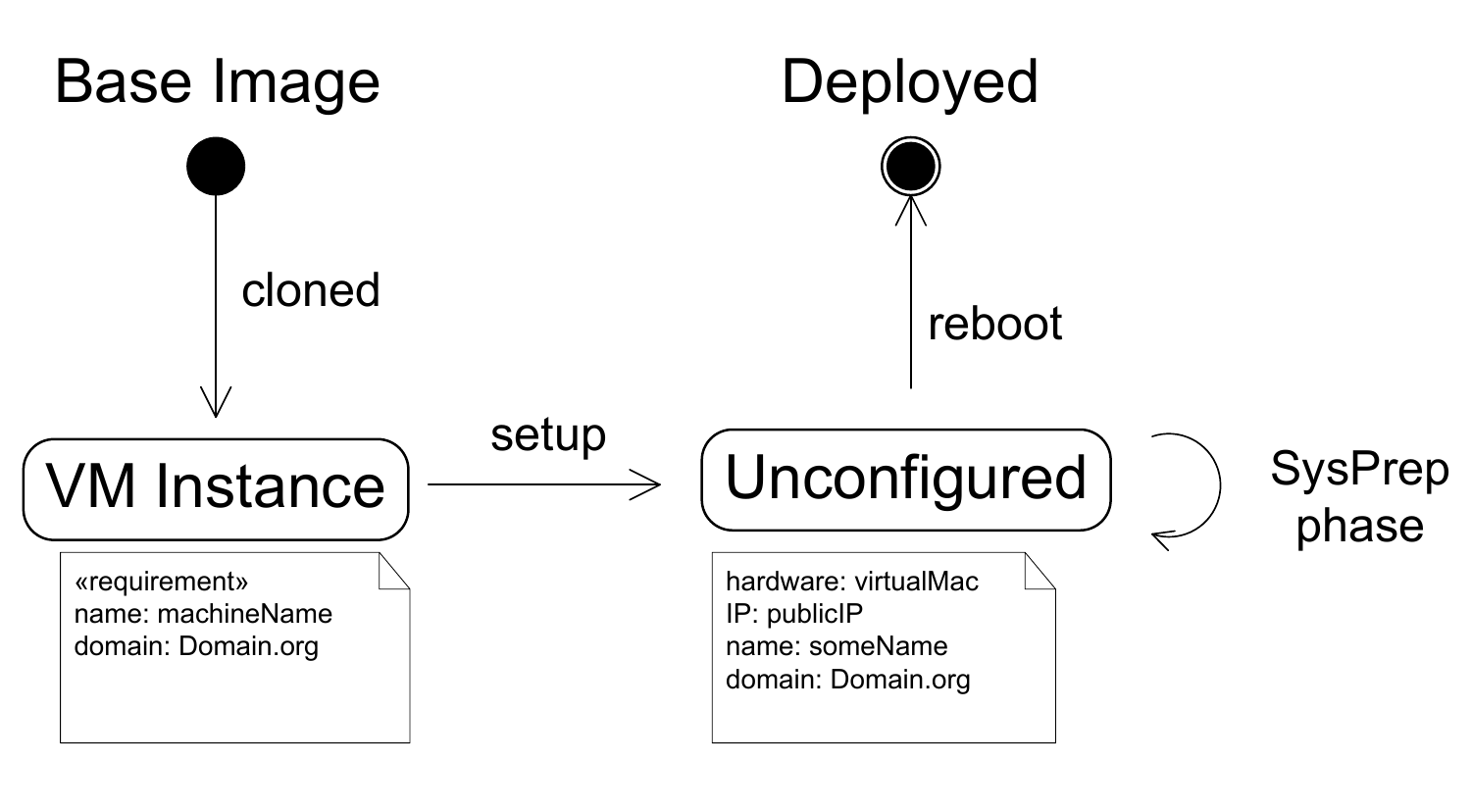}
   \caption{State diagram of a VM starting with the base image, and the stages it goes through to get to running state.}
   \label{sysprep}
\end{figure}

For Windows VM, as shown in Fig.  \ref{sysprep}, before the reboot there is an additional stage of reseting virtual machine's security ID as required by the domain controller and is linked to the active directory entry. If the same security ID is used for all windows VM's, then they can't join the domain. This process is delegated to Microsoft's System Preparation (SysPrep) tool which configures the VM using a local configuration file. Once this configuration is completed; the machine is rebooted and then is available to end-user via the remote desktop connection.

\section{RESULTS}

All four different architectures were evaluated for four different deployment scenarios. Each scenario was run three times and the results were averaged and are presented in this section. Any computing infrastructure when used by multiple users goes under different cycles of demand which results in reduced supply of available resources on the infrastructure to deliver optimal service quality.

We were particularly interested in following deployment scenarios where 10 virtual machines were deployed:
\begin {itemize}
\item Single Burst (SB): All virtual machines are sent in a burst mode but restricted to one server only. This is the most resource-intensive request.
\item Multi Burst (MB): All virtual machines were sent in a burst mode to multiple servers.
\item Single Interval (SI): All virtual machines were sent after an interval of 3 mins to one server only.
\item Multi Interval (MI): All virtual machines were sent after an interval of 3 mins to multiple servers. This is the least resource-intensive request.
\end{itemize}

\begin{figure}[htb]
   \centering
   \includegraphics*[width=85mm]{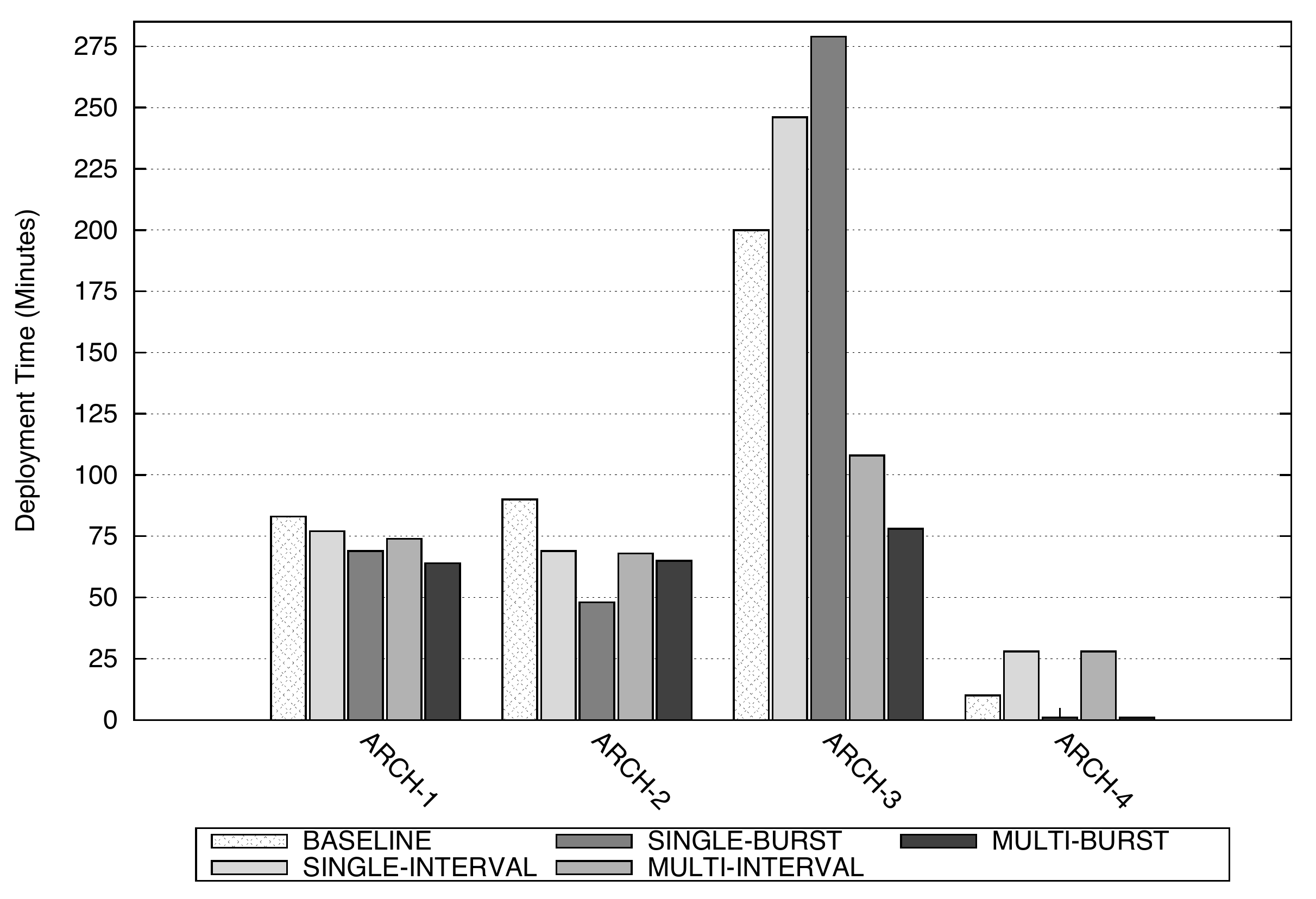}
   \caption{Aggregated total time taken to deploy all VM's in each test configuration for all architectures.}
   \label{totaltime}
\end{figure}

The overall deployment times, as shown in Fig. \ref{totaltime}, for \textit{$arch_1$} and \textit{$arch_2$} are very close to each other for all four test configuration. Both of these architectures were using a NFS based shared storage where all VM images were cloned on a storage server, and only VM setup commands were sent to the ESXi nodes with image pointers. All ten VM's got deployed within 75 mins of request initialization. This is the lower bound of the system.  Where as \textit{$arch_3$} is using distributed storage but every time a VM request is made, a new image is transferred over SSH which is a very slow process and can take up more then 200 mins for some configurations. \textit{$arch_4$} is most interesting that clearly shows the optimization we implemented for auto-deployment of VM images prior to requests in the background which results in VM being deployed within 10mins.

\begin{figure}[htb]
   \centering
   \includegraphics*[width=85mm]{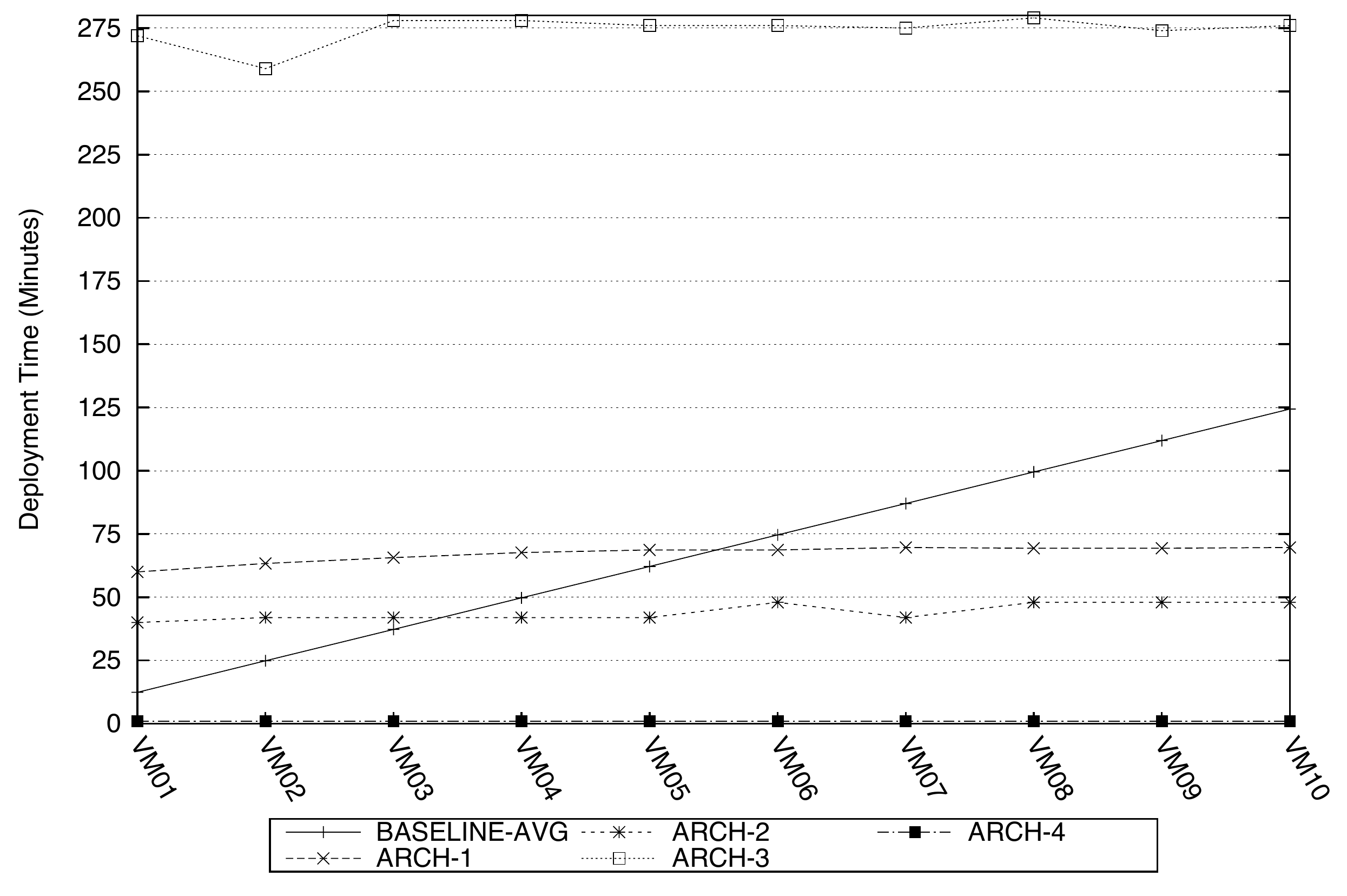}
   \caption{Evolution of deployment time for single-burst (SB) configuration.}
   \label{wait_sb}
\end{figure}

\begin{figure}[htb]
   \centering
   \includegraphics*[width=85mm]{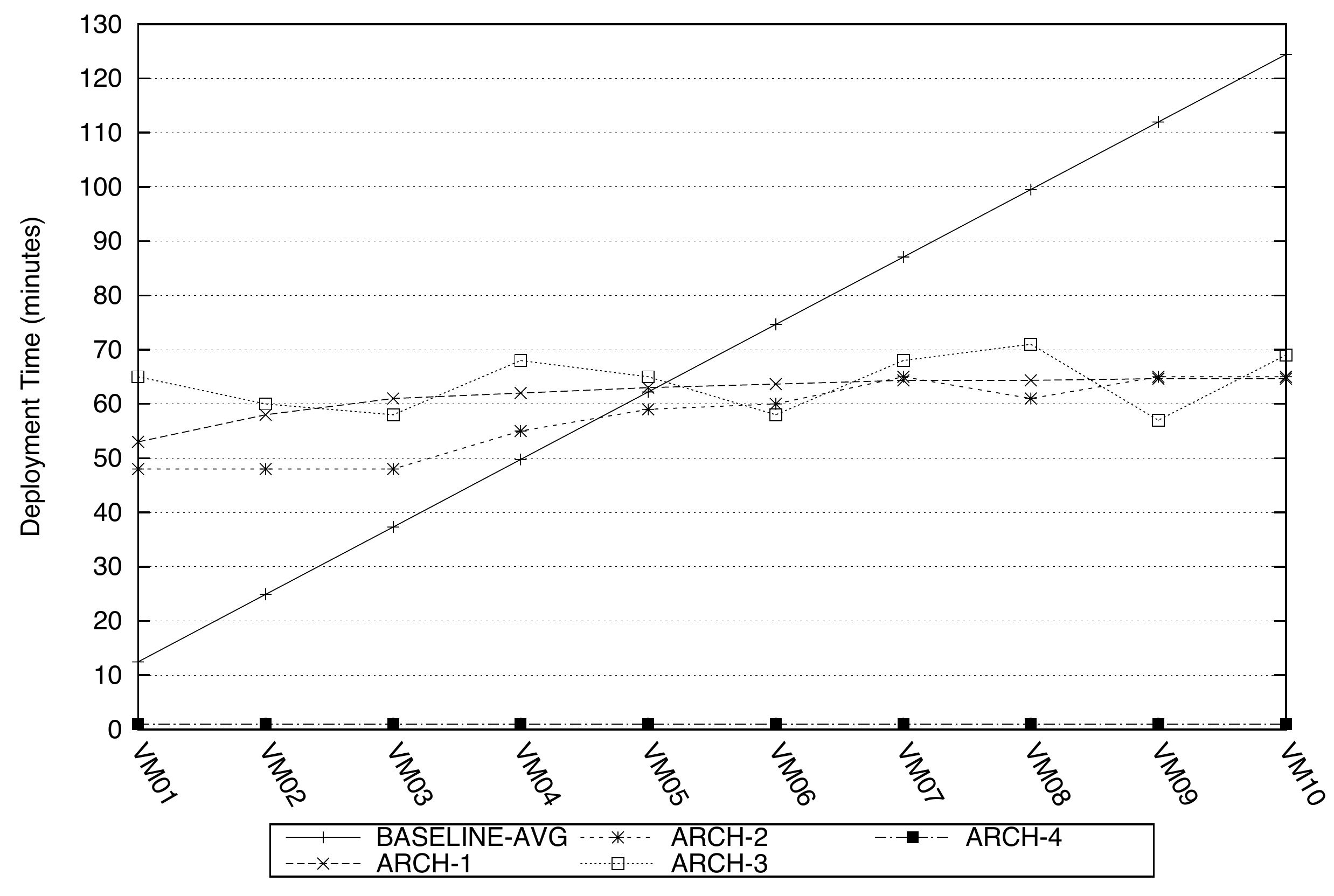}
   \caption{Evolution of deployment time for multi-burst (MB) configuration.}
   \label{wait_mb}
\end{figure}

\begin{figure}[htb]
   \centering
   \includegraphics*[width=85mm]{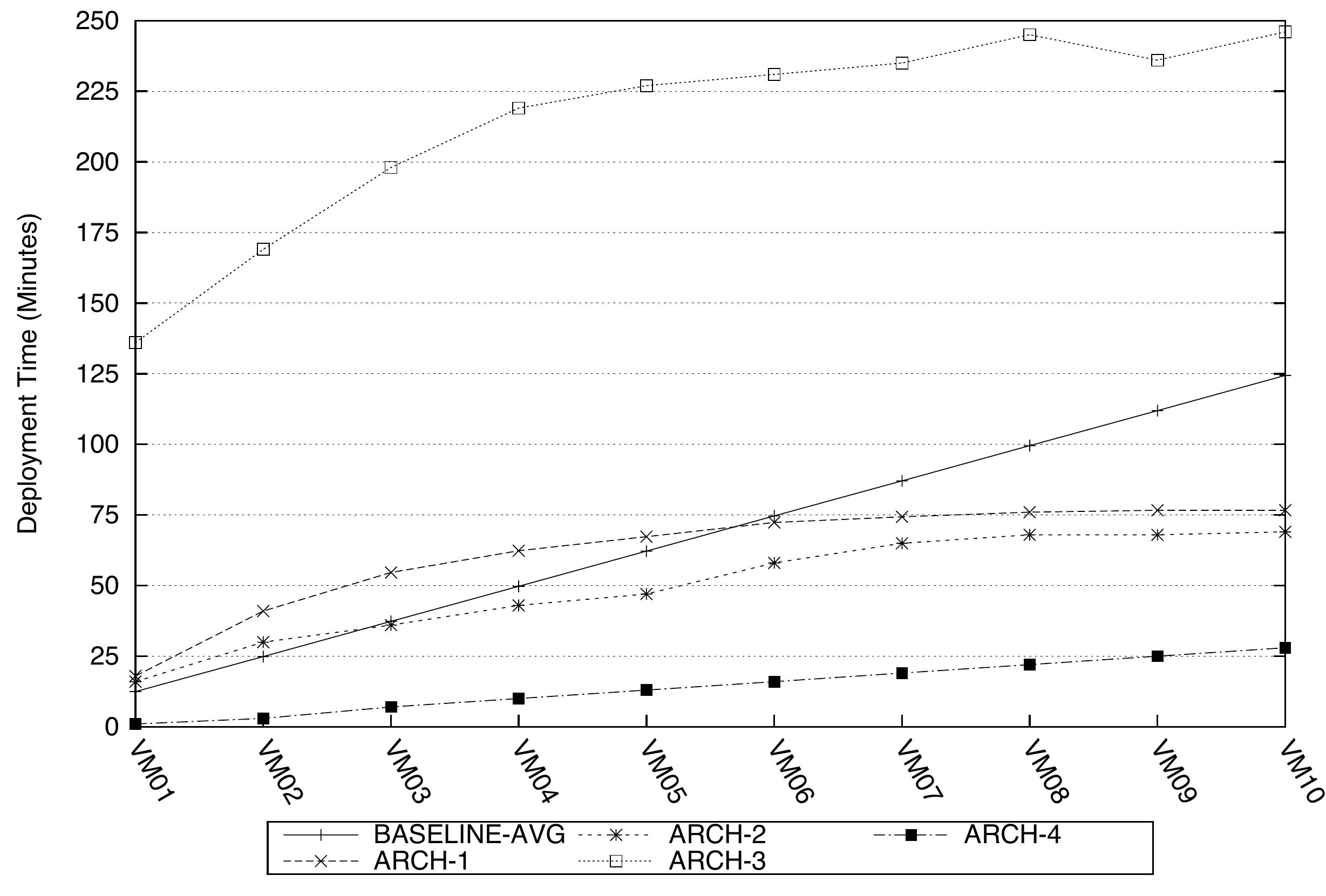}
   \caption{Evolution of deployment time for single-interval (SI) configuration.}
   \label{wait_si}
\end{figure}

Figure \ref{wait_sb} shows that cumulative wait time for VM deployment is fairly stable for \textit{$arch_1$} and \textit{$arch_2$}. \textit{$arch_3$} is taking the highest amount of time and \textit{$arch_4$} least. Similar pattern of wait time is observable in Fig. \ref{wait_mb} and \ref{wait_si} as well. For MI configuration which is least resource intensive, \textit{$arch_2$} and \textit{$arch_3$} have a similar wait time which shows that both architectures are suitable for small-scale cloud deployments that a VM gets deployed within 30 mins. The only issue is central storage being a single-point of failure incase a hardware fault occurs. \textit{$arch_4$} keeps on out perfoming all other configuration, and hence was selected as the target deployment on our private cloud.


\begin{figure}[htb]
   \centering
   \includegraphics*[width=85mm]{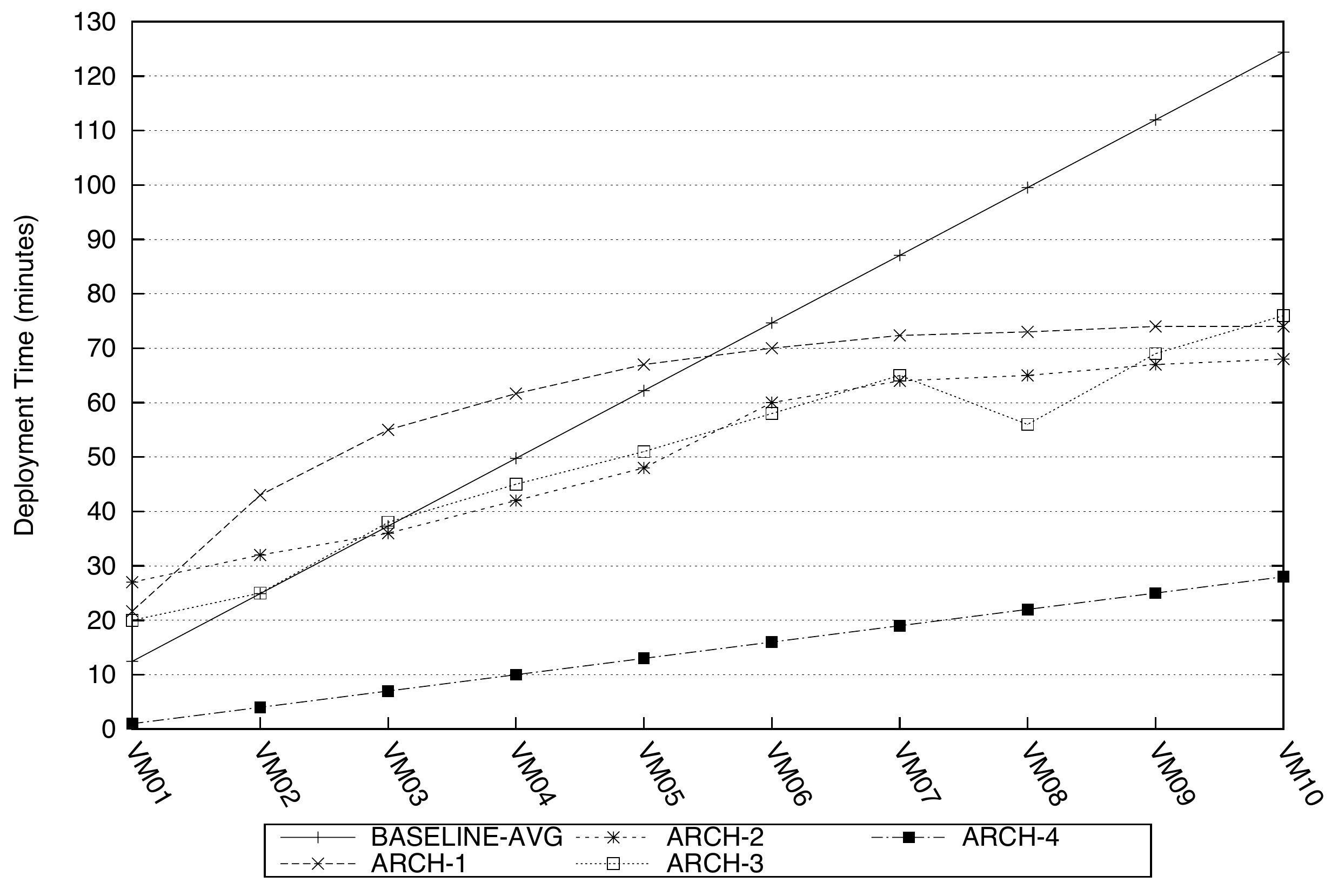}
   \caption{Evolution of deployment time for multi-interval (MI) configuration.}
   \label{wait_mi}
\end{figure}



\section{CONCLUSION}

Virtualization technology and the emerging cloud computing platform provides innovative way to utilize physical infrastructure in an elastic and flexible manner. It enables system administrators to meet various cycles of infrastructure demand when multiple user communities access the same hardware. Without cloud computing, it requires automated system administration tools to provide uniform access to storage, CPU, network and memory of a cluster of machines. Whereas, present day cloud computing management systems allows system administrators to not only virtualize their infrastructure which enables to deploy more applications/machines (virtual) on the shared physical hardware but also reduces the application deployment lifecycle.
 
In this paper, we have attempted to evaluate small scale private cloud infrastructure (up to 10 hardware servers) for software development teams so that they could quickly and on-demand request machine resources using available virtual machine technology. Our study have  highlighted that for optimal performance; a high-performance shared storage SAN and high-network bandwidth is preferable but this is often not possible for small scale deployments due to financial constraints. The experiments conducted in this study have indicated that a small scale private cloud is a feasible option without costly SAN or high-speed networking gear. 

The results have also shown that distributed storage using locally cached images when managed using a centralized cloud platform (in our study we used OpenNebula Pro) is a practical option to setup local clouds where users can setup their virtual machines on demand within 15mins (from request to machine boot up) while keeping the cost of the underlying infrastructure low.

\end{document}